\documentclass[
reprint,
nofootinbib,
 amsmath,amssymb,
 aps, superscriptaddress,
]{revtex4-2}

\usepackage{graphicx}
\usepackage{dcolumn}
\usepackage{bm}

\usepackage{hyperref}

\usepackage[compat=1.1.0]{tikz-feynman}
\tikzfeynmanset{warn luatex=false}
\usetikzlibrary{positioning,arrows}
\usetikzlibrary{decorations.pathmorphing}
\usetikzlibrary{decorations.markings}
\usetikzlibrary{shapes,shadows}

\usepackage{subfiles}
\usepackage{color}
\usepackage{slashed}
\usepackage{booktabs}
\usepackage{tabularx}
\usepackage{makecell}
\usepackage[normalem]{ulem}
\usepackage[svgnames]{xcolor}
\usepackage{subcaption}

\def\gm{\gamma}

\def\ep{\epsilon}
\def\epm1{\frac{1}{\epsilon}}
\def\epm2{\frac{1}{\epsilon^{2}}}
\def\epm3{\frac{1}{\epsilon^{3}}}
\def\epm4{\frac{1}{\epsilon^{4}}}

\def\rnM{{\cal M}}
\def\bt{\beta}

\def\bt0{\beta_{0}}
\def\bt1{\beta_{1}}
\def\bt2{\beta_{2}}
\def\bt3{\beta_{3}}
\def\gm0{\gamma_{0}}
\def\gm1{\gamma_{1}}
\def\gm2{\gamma_{2}}
\def\gm3{\gamma_{3}}

\def\l{\left}
\def\r{\right}

\def\smu{\left( \frac{s}{\mu^2} \right)}

\def\I{{\cal I}}

\newcommand{\abiss}{{\sc Abiss~}}
\newcommand{\seasyde}{{\sc SeaSyde~}}
\newcommand{\feynarts}{{\sc FeynArts~}}
\newcommand{\kira}{{\sc Kira~}}

\newcommand{\litered}{{\sc LiteRed~}}
\newcommand{\mathematica}{{\sc Mathematica~}}
\newcommand{\AMFlow}{{\sc AMFlow}~}

\newcommand{\be}{\begin{equation}}
\newcommand{\ee}{\end{equation}}
\newcommand{\bea}{\begin{eqnarray}}
\newcommand{\eea}{\end{eqnarray}}

\newcommand{\oaas}{${\cal O}(\alpha\alpha_s)~$}

\begin{document}

\preprint{APS/123-QED}

\title{Two-loop mixed QCD-EW corrections to bottom-pair production\\ in electron-positron collisions}

\author{Simone Devoto}
\email{simone.devoto@ugent.be}
\affiliation{Department of Physics and Astronomy, Ghent University, 9000 Ghent, Belgium}

\author{Chinmoy Dey}
\email{chinmoy@sjtu.edu.cn}
\affiliation{Tsung-Dao Lee Institute and  School of Physics and Astronomy, Shanghai Jiao Tong University,
800 Lisuo Road, Shanghai, 200240 China}
\affiliation{Shanghai Key Laboratory for Particle Physics and Cosmology, 
Key Laboratory for Particle Astrophysics and Cosmology (MOE), 
Shanghai Jiao Tong University, Shanghai 200240, China}

\author{Michele Dradi}
\email{michele.dradi@physik.uzh.ch}
\affiliation{Physics department ``Aldo Pontremoli'',
  University of Milan and INFN, \\ Milan Unit, I-20133 Milan, Italy}
\affiliation{Department of Physics,
  University of Zurich, Zurich, Switzerland}
  
\author{Justin T.~Fagnoni}
\email{jfagnoni@umass.edu}
\affiliation{Amherst Center for Fundamental Interactions, Department of Physics, University of Massachusetts, Amherst, MA 01003, USA}

\author{Ziyang Huang}
\email{slybgodh@sjtu.edu.cn}
\affiliation{Tsung-Dao Lee Institute and  School of Physics and Astronomy, Shanghai Jiao Tong University,
800 Lisuo Road, Shanghai, 200240 China}
\affiliation{Shanghai Key Laboratory for Particle Physics and Cosmology, 
Key Laboratory for Particle Astrophysics and Cosmology (MOE), 
Shanghai Jiao Tong University, Shanghai 200240, China}

\author{Xinchen Li}
\email{starleetdli@gmail.com}
\affiliation{Tsung-Dao Lee Institute and  School of Physics and Astronomy, Shanghai Jiao Tong University,
800 Lisuo Road, Shanghai, 200240 China}
\affiliation{Shanghai Key Laboratory for Particle Physics and Cosmology, 
Key Laboratory for Particle Astrophysics and Cosmology (MOE), 
Shanghai Jiao Tong University, Shanghai 200240, China}

\author{Michael J.~Ramsey-Musolf}
\email{mjrm@sjtu.edu.cn,mjrm@caltech.edu}
\affiliation{Tsung-Dao Lee Institute and  School of Physics and Astronomy, Shanghai Jiao Tong University,
800 Lisuo Road, Shanghai, 200240 China}
\affiliation{Shanghai Key Laboratory for Particle Physics and Cosmology, 
Key Laboratory for Particle Astrophysics and Cosmology (MOE), 
Shanghai Jiao Tong University, Shanghai 200240, China}
\affiliation{Kellogg Radiation Laboratory, California Institute of Technology, Pasadena, CA 91125, USA}
  
\author{Alessandro Vicini}
\email{alessandro.vicini@unimi.it}
\affiliation{Physics department ``Aldo Pontremoli'',
  University of Milan and INFN, \\ Milan Unit, I-20133 Milan, Italy}

\author{Jiang Zhu}
\email{jackpotzhujiang@gmail.com}
\affiliation{Tsung-Dao Lee Institute and  School of Physics and Astronomy, Shanghai Jiao Tong University,
800 Lisuo Road, Shanghai, 200240 China}
\affiliation{Shanghai Key Laboratory for Particle Physics and Cosmology, 
Key Laboratory for Particle Astrophysics and Cosmology (MOE), 
Shanghai Jiao Tong University, Shanghai 200240, China}

\date{\today}

\begin{abstract}
  We present the two-loop mixed strong-electroweak virtual corrections
  to the production of a massive bottom-antibottom quark pair in electron-positron collisions. 
 The evaluation of all the relevant Feynman integrals,
 including those with up to five internal massive lines with three different mass values, has been worked out relying on semi-analytical techniques,
 using complex-valued masses for the electroweak gauge bosons.
 The ultraviolet-renormalized and infrared-subtracted finite correction contributes to the NNLO QCD-EW cross section of the inclusive production of a bottom-quark pair.
The methodology adopted in this calculation can be extended to the study of fully electroweak second-order corrections to benchmark precision calculations relevant to both the high-luminosity phase of the Large Hadron Collider and prospective future lepton colliders.

\end{abstract}

\maketitle

\section{Introduction}
The high-luminosity operation phase of the CERN Large Hadron Collider (HL-LHC)~\cite{Azzi:2019yne} and the precision achievable at a future lepton collider such as the CEPC, the FCC-ee, or a Linear Collider~\cite{FCC:2018byv,CEPCStudyGroup:2018ghi,ILC:2013jhg}  set unprecedented challenges in the evaluation of theoretical predictions for cross sections and asymmetries, the observables measured in high-energy scattering processes.
Prospects for the measurement of cross sections with sub-percent experimental precision at the highest energy scales available at hadron colliders~\cite{Armadillo:2026wdb}, namely with invariant masses of several TeV, requires the availability not only of QCD results, but also of mixed QCD-electroweak and purely electroweak (EW) corrections at second order in perturbation theory. More precise theoretical results will improve the statistical significance of the comparison with the data and strengthen the evidence of a deviation, if any, of the data from the Standard Model (SM) predictions.

In a complementary context, the very high luminosity which will be collected at future lepton colliders at the $Z$-boson resonance will push the theoretical needs towards the completion of the calculation of the third-order electroweak corrections to the production of a fermion pair~\cite{Blondel:2019qlh,Ge:2024pfn}.
The high-intensity frontier explored at B-factories such as Belle II~\cite{Belle-II:2022cgf} or in the study of Møller scattering~\cite{Demiroglu:2024wys,Chen:2026qcq} will demand the availability of the complete second-order electroweak corrections to the relevant processes, as well.
The development of new methodologies for evaluating higher-order perturbative corrections in both the strong and electroweak interactions is thus urgently needed, in order to reduce the theoretical uncertainties down to the level of the experimental ones.

The evaluation of SM scattering amplitudes has seen tremendous progress in the last 25 years, with an increasing level of automation of many calculational steps and improved systematic understanding of the mathematical properties of the Feynman integrals entering computations of the quantum virtual corrections. 
Automated evaluation of physical cross sections at  next-to-leading-order (NLO) accuracy has become  standard~\cite{Alwall:2014hca,Alioli:2010xd,Cascioli:2011va}. The evaluation of two-loop Feynman integrals has become a systematic procedure~\cite{Gehrmann:1999as,Henn:2013pwa,Liu:2022chg}, allowing the completion of several calculations with next-to-next-to-leading-order (NNLO) accuracy. 
The evaluation of realistic fiducial cross sections has become possible with the automation of subtraction procedures in various public tools~\cite{Grazzini:2017mhc,NNLOJET:2025rno}.

Despite  these advances, the computation of multiscale virtual amplitudes remains an open frontier. At present, it is the least standardized part of the general workflow.
The complexity of a multi-loop scattering amplitude can be described by the number of energy scales present in the process, including the kinematical invariants, the virtuality of the external legs, and the internal particle masses.  
Not only do these parameters lead to Feynman integrals with a more involved analytic structure, but the presence of a large number of distinct symbols, which label the different masses and kinematical invariants, leads in turn to a fast growth in the size of the rational functions multiplying the Feynman integrals, easily reaching the GB level for individual coefficients present in the amplitude. 

Considering different combinations of the number of loops and external particles, we find examples of current computational frontiers: vertex form factors with external massive particles at three-loop level in QCD \cite{Blumlein:2019oas,Datta:2023otd}; $2\to 2$ scattering amplitudes of massless particles in QED at three-loop level \cite{Crisanti:2026luo};  scattering amplitudes of $2\to 3$ particles, at two-loop level in QCD, with one external mass \cite{Abreu:2023rco}; and  scattering amplitudes of $2\to 2$ massless particles, with up to two distinct internal massive lines, as found in the QCD-EW subsets of two-loop corrections \cite{Armadillo:2022bgm,Buccioni:2022kgy,Armadillo:2024nwk}. 
The presence of several internal and external massive particles, like in the case of double Higgs production, can be treated with expansions and numerical evaluation of the Feynman integrals \cite{Borowka:2016ehy}. 

In this paper, we push the application of the available tools towards the level of complexity expected in the future evaluation of the complete second-order EW corrections to $2\to 2$ scattering.
The main goal of this work is to develop an algorithm for calculating the effects due to two-loop virtual corrections in the presence of multiple external and internal mass scales, including ultraviolet (UV) renormalization and the handling of infrared (IR) divergences.
As a test of our approach we consider the mixed QCD-EW corrections to the production of a pair of massive bottom quarks in electron-positron annihilation. 
This production process has been studied at LEP/SLD with a center-of-mass energy $\sqrt{s}=M_{Z}$, the $Z$-boson mass. 
The measured forward-backward asymmetry of this final state has allowed a precise determination of the effective weak-mixing angle, which is however in tension with the best SM prediction \cite{Awramik:2006uz,Dubovyk:2016aqv} (see also \cite{Degrassi:1993ij,Chetyrkin:1993jp,Chetyrkin:1993yp}) and with the value extracted from the study of the left-right polarization asymmetry $A_{LR}$ \cite{ALEPH:2005ab}.
Since the extraction of $\sin^2\theta_W^\text{eff}$ from the data requires a sophisticated analysis and input from the best available theoretical cross sections, it is well worth assessing the impact of mixed QCD-EW effects.
A complete phenomenological discussion requires the availability of the inclusive cross section $\sigma(e^+e^-\to b \bar b +X)$, and the bottleneck is found in the evaluation of the two-loop virtual corrections. 

As the main result of this paper, we present the calculation of the two-loop amplitudes for the mixed QCD-EW corrections to bottom-pair production at lepton colliders, retaining the exact dependence on the bottom-quark mass. This work:
\begin{itemize}
\item Describes the cancellation of all the UV and IR poles appearing at the considered order;
\item Demonstrates the independence of the results from the conventions adopted to implement Kreimer's treatment of chiral couplings in $d=4-2\epsilon$ space-time dimensions;
\item Retains the exact dependence on the non-vanishing external-quark masses;
\item Achieves exact solutions of the Feynman integrals with up to five internal massive lines and three distinct mass values.
\end{itemize}
This computation and the corresponding set of technical developments constitutes a jump in complexity compared to the existing results for neutral- and charged-current Drell-Yan \cite{Armadillo:2022bgm,Armadillo:2024nwk}, 
where the final-state particle masses have been retained only where necessary to regularize collinear singularities.

In the following sections, we describe our workflow for computing the UV-renormalized and IR-subtracted interference term between the tree-level and mixed QCD-EW two-loop amplitudes of the process $e^+e^-\to b\bar b$. In Section \ref{method}, we define the process and its kinematics and we describe our workflow.  In Section \ref{UV_IR}, we discuss the details of UV renormalization and IR subtraction.  In Section \ref{amp}, we describe the physical and mathematical content of the amplitude. In Section \ref{cross-checks}, we discuss our validation techniques. In Section \ref{results}, we present our benchmark results. Finally, in Section \ref{conclusions}, we draw our conclusions. 
\section{Framework of the calculation} \label{method}
\subsection{The process}
In this paper, we discuss the
two-loop ${\cal O}(\alpha^3\alpha_s)$ contributions to the squared amplitude for the production of a pair of massive bottom quarks in electron-positron annihilation, 
\begin{equation}
 e^-(p_1) + e^+(p_2) \rightarrow b(p_3) + \bar{b} (p_4),
\label{eq:process}
\end{equation} 
which are a necessary component of the total inclusive cross section. 
We consider massive final-state quarks, while the initial-state leptons are considered massless:
\begin{align}
    &p_1^2=p^2_2=0\;,&p_3^2=p_4^2=m_b^2.
\end{align}
The square of the bare amplitude $|\rnM_0 \rangle$ can be expanded in powers of the  weak ($\alpha_0$) and strong ($\alpha_{s,0}$) bare couplings as follows:
\begin{equation}
\begin{split}
&\langle \rnM_0|\rnM_0\rangle = 16\pi^2 \alpha^2_0 \\
&\times \sum_{j,k,l,m=0}^\infty \left( \frac{\alpha_{s,0}}{4\pi}\right)^{j+l} \left(\frac{\alpha_0}{4\pi}\right)^{k+m}\langle\rnM^{(l,m)}_0 | \rnM^{(j,k)}_0\rangle.
\label{amp_sq}
\end{split}
\end{equation}
Here, the superscripts $(j,k)$ are used to denote corrections that are $j^\text{th}$ order in $\alpha_s$ and $k^\text{th}$ order in $\alpha$.
Up to complex conjugation, there are two contributions at ${\cal O}(\alpha^3 \alpha_{s})$: $\langle \rnM^{(1,0)}_0|\rnM^{(0,1)}_0\rangle$ and $\langle \rnM^{(0,0)}_0| \rnM^{(1,1)}_0\rangle$. We consider the latter matrix element in this work. 

The bare matrix element $\langle \rnM^{(0,0)}_0| \rnM^{(1,1)}_0\rangle$ contains poles that are of both UV and IR origin. The corresponding finite matrix element $\langle \rnM^{(0,0)}| \rnM^{(1,1),fin}\rangle$ is obtained through UV renormalization and IR subtraction, which are discussed in Sections \ref{sec:UV} and \ref{sec:IR}, respectively. This quantity enters the definition of the hard-scattering function $H^{(1,1)}$ \cite{Armadillo:2022bgm}:
\begin{equation}
    H^{(1,1)} = \frac{1}{16}\left[2~\text{Re} \left(\frac{\langle \rnM^{(0,0)}| \rnM^{(1,1),fin}\rangle}{\langle \rnM^{(0,0)}|  \rnM^{(0,0)}\rangle} \right)\right].\label{H11}
\end{equation}
The hard function captures the short-distance process-dependent contributions to the total inclusive cross section that share the Born-level $2\rightarrow 2$ kinematics. These are distinct from the real-emission contributions with $2 \rightarrow 2 + n$ kinematics, where $n$ is the number of emitted gluons and/or photons.

\subsection{Methodology}
\label{sec:metho}

\begin{figure}[htbp]     
    \centering
    \includegraphics[width=0.49\textwidth]{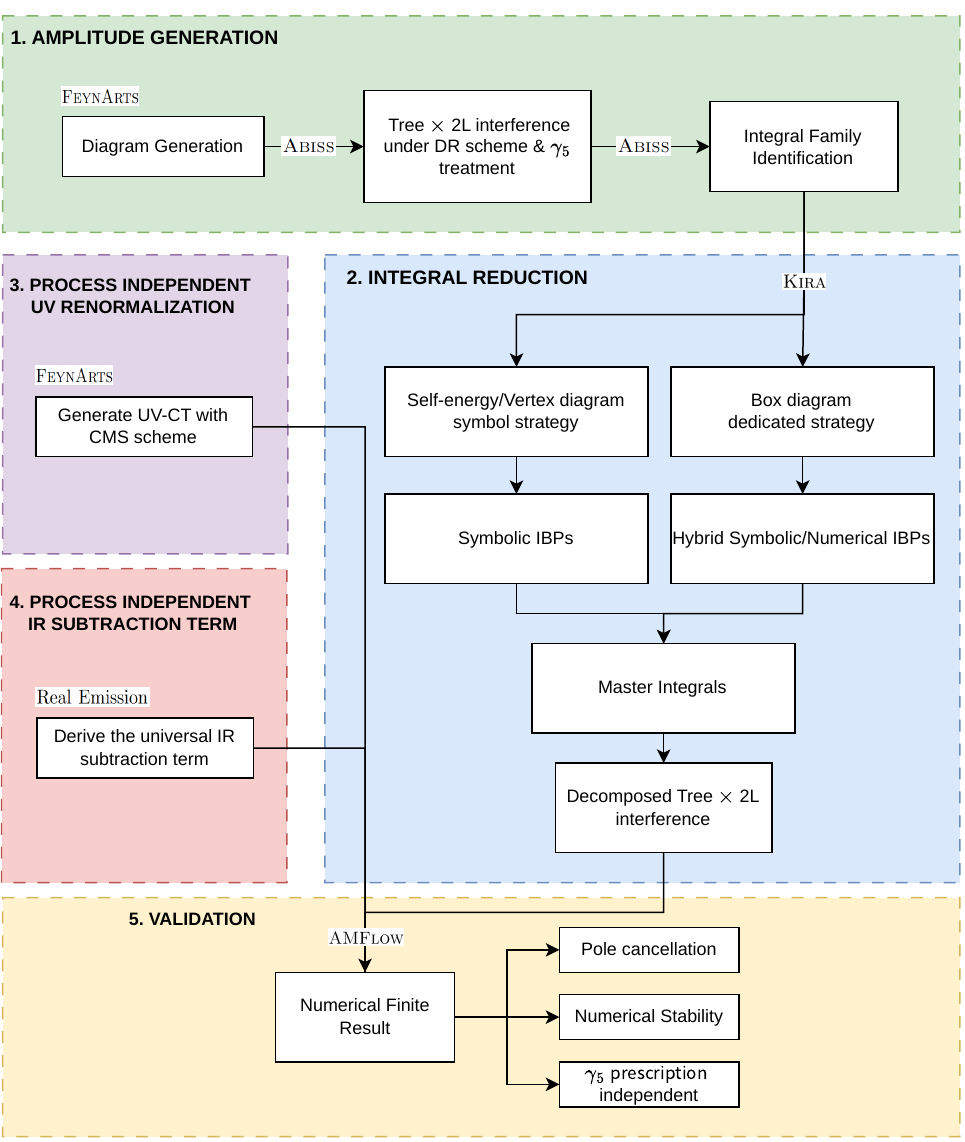}
   
    \caption{Illustration of all the different steps of the workflow employed in our calculation.}
     \label{workflow}
\end{figure}

In this section, we summarize the methodological framework of our calculation, showing in detail how we handle the algebraic manipulation in the evaluation of interference terms, how we circumvent the problems connected to the size and stability of the partial results, and how we perform the final numerical evaluation of the two-loop amplitude.  We show in Figure \ref{workflow} the various elements of this framework and the corresponding workflow.

\begin{enumerate}
    \item All Feynman diagrams contributing to the process are generated with \feynarts \cite{Hahn:2000kx}, using custom-made model files extending the complex-mass scheme to second-order calculations.
    The interference between the two-loop virtual corrections and the tree-level amplitude is evaluated with the \mathematica package \abiss \cite{abiss}, summing over the polarizations of the external fermion fields.
    The calculation is performed in dimensional regularization, with $d=4-2\epsilon$ being the number of space-time dimensions.
 The Dirac matrix $\gamma_5$, which appears in the SM chiral couplings, is treated following Kreimer's prescription, in which the fully antisymmetric tensor $\epsilon^{\mu\nu\alpha\beta}$ arising from Dirac traces is regarded as a purely 4-dimensional object.
This procedure is described in detail in \cite{Heller:2020owb,Armadillo:2026wdb,abiss}.
The identification of the Feynman integrals and their labeling in terms of integral families is automatically realized within {\sc{Abiss}}, which delivers the expressions of the interference terms as combinations of the scalar integrals contributing to the final result.
    
\item An external code, in our case \kira \cite{Maierhoefer:2017hyi}, is used to find all the relevant integration-by-parts (IBP) 
reduction rules needed to express any scalar Feynman integral as a linear combination of master integrals (MIs), the basic building blocks for the amplitude representation. 
Whenever possible, we aim for a fully symbolic solution, which guarantees an improved flexibility for the later usage of our results. The application within \kira of polynomial reconstruction algorithms, such as the reconstruction over finite fields implemented with {\sc FireFly}\cite{Klappert:2020aqs}, is able to reduce both the necessary evaluation time and the final size of the replacement rules. 
    
In the case of the box corrections (see Figure~\ref{fig3c}), even with state-of-the-art tools we have not been able to obtain a fully symbolic reduction to MIs.
To obtain the reduction rules, we set to numerical values the mass of all the internal particles, with the exception of the electroweak bosons, thus keeping only the analytic dependence on the $Z$- and $W$-boson masses and the Mandelstam invariants.
Furthermore, we make sure, at each step of the procedure, to generate the minimal set of reduction rules required to cover the subspace spanned by our target MIs, splitting it, when necessary, into appropriate  subsectors. 
    
\item The UV renormalization of the amplitude is performed after the evaluation of all the relevant counterterms for the masses, couplings, and fields involved in the scattering, including corrections of \oaas \cite{Dittmaier:2020vra,Armadillo:2022bgm}. We employ the background-field method (BFM) \cite{Denner:1994xt, Denner:2019vbn} to evaluate all contributions to the renormalized amplitude, which provides avenues for cross-checks of our results (see Section \ref{sec:UV} for further discussion).
The counterterms are computed in the on-shell scheme and arranged in a library, ready for any other calculation at this perturbative order. To perform numerical studies, the counterterms have been evaluated using the same input parameters as the process under discussion, and have been written as Laurent expansions in the dimensional regulator $\epsilon$. 
The Feynman diagrams featuring the insertion of UV counterterms have also been generated with \feynarts and have been summed with the rest of the amplitude following the same procedure outlined above.
    
\item The subtraction of the soft and soft-collinear IR divergences appearing in the virtual amplitude relies on their universality, which allows in turn their factorization with respect to the underlying hard process. NLO and NNLO subtraction operators have been derived, along the lines of \cite{deFlorian:2018wcj}, starting from the general structure of the IR singularities in QCD \cite{Ferroglia:2009ii} (see Section~\ref{sec:IR}).
We compute the subtracted two-loop amplitude by combining (1) the renormalized, unsubtracted contribution with the products of the NNLO subtraction operator and  the tree-level amplitude, and (2) the NLO subtraction operators and the UV-renormalized one-loop IR-subtracted amplitudes.
    
\item As a final step, the values of the identified set of MIs are computed with \AMFlow \cite{Liu:2022chg,Huang:2026rjb}, and several checks are performed to establish the reliability of both the calculation and the overall computational workflow. These include (1) explicitly verifying the cancellation of all divergent terms; (2) assessing the numerical stability of the finite remainder; and (3) confirming the independence of the results from the conventions adopted in Kreimer's prescription for evaluating fermionic traces involving the $\gamma_5$ Dirac matrix.
\end{enumerate}

\section{UV renormalization and IR subtraction at \texorpdfstring{${\cal O}(\alpha\alpha_s)$}{O(alpha alphaS)}} \label{UV_IR}
\subsection{Ultraviolet renormalization}
\label{sec:UV}

The renormalized amplitude $|\rnM\rangle$ is obtained by replacing the bare parameters and fields in $|\rnM_0\rangle$ with their renormalized counterparts. This can be illustrated schematically as follows: 
\begin{equation}
\begin{split}
&|\rnM
\rangle=Z_{e^-}^\frac12 Z_{e^+}^\frac12 Z_b^\frac12 Z_{\bar b}^\frac12\\
&\phantom{|\rnM
\rangle=} \times 16\pi^2  \sum_{j,k=0}^\infty \left(\frac{g + \delta g_s}{4\pi} \right)^{2j} \left(\frac{e + \delta e}{4\pi} \right)^{2k+2} \\
&\phantom{|\rnM
\rangle=}  \times| \rnM_0^{(j,k)}(s,t,\{m_{f}+\delta m_f\}, \{\mu_{B}^2+\delta\mu_B^2\}) \rangle\\
&= 4\pi\alpha \sum_{j,k=0}^\infty \left(\frac{\alpha_s}{4\pi}\right)^j\left(\frac{\alpha}{4\pi}\right)^{k}|\rnM^{(j,k)}(s,t,\{m_f\},\{\mu_B^2\})\rangle,\label{ren_amp}
\end{split}
\end{equation} 
where $s$ and $t$ are Mandelstam variables, $g_s=\sqrt{4\pi\alpha_s}$ and $e=\sqrt{4\pi\alpha}$ are the renormalized strong and electromagnetic couplings, $\{m_f\}$ is the set of renormalized masses of the fermions, $Z_f^\frac12$ is the wave-function renormalization factor of fermion $f$, and $\{\mu_B\}$ is the set of renormalized masses of the gauge bosons. The quantities $\delta g_s$, $\delta e$, $\delta m_f$, and $\delta \mu_B^2$ represent the counterterms associated with the couplings and masses.
Note that in Equation \eqref{ren_amp}, only the external fields receive wave-function renormalization factors, as those associated with internal fields are guaranteed to cancel between renormalized vertices and propagators. At ${\cal O}(\alpha \alpha_s)$, the external electrons receive at most one-loop corrections, whereas the external bottom quarks require full two-loop renormalization.


We evaluate all the contributions to the renormalized amplitude $|\rnM^{(1,1)}\rangle$ using the BFM \cite{Denner:1994xt, Denner:2019vbn}, in which the scalar and vector fields of the classical Lagrangian are split into the sum of background and quantum fields, with only the latter taken as integration variables in the SM generating functional. 
Within this formalism, the gauge invariance of the classical SM Lagrangian leads to QED-like Ward identities for the 1-particle irreducible (1PI) vertex functions involving external background photon fields. 
Simple Ward identities are also satisfied by the 1PI vertex functions involving massive background gauge fields. In particular, the 1PI massive gauge boson self-energies are connected with the two-point functions of the associated pseudo-Goldstone bosons and tadpole contributions, which we organize following the parameter-renormalized tadpole scheme~\cite{Sirlin:1980nh, Sirlin:1985ux,Denner:2019vbn}.

The relations that hold in the BFM greatly simplify the cancellation pattern of the UV singularities both at the one-loop level and at $\mathcal{O}(\alpha \alpha_s)$. For instance, the $\epsilon^{-k}$ poles in the combination of gauge-boson self-energy and mass counterterm contributions exactly cancel with the charge counterterms $\delta e$ and $\delta s_W^2$ (where $s_W^2 \equiv \sin^2\theta_W = 1 - \mu_W^2/\mu_Z^2$) inserted at the adjacent vertices, as described in \cite{Armadillo:2022bgm}. As a consequence, the initial- and final-state vertex corrections are rendered UV finite by the insertion of wave-function counterterms on the external fermion lines, which is not generally true in the conventional method of quantization.
These relations, which are guaranteed to hold by the symmetries of the effective action, provide additional crosschecks of our numerical results.

The counterterms are evaluated in the complex-mass scheme (CMS) \cite{Denner:2005fg}, where the square of the gauge-boson masses $\mu_W^2$ and $\mu_Z^2$ are taken as the complex poles of the $W$ and $Z$ propagators: 
\be
\mu_V^2=M_V^2 - i M_V \Gamma_V, \ \ \ \ V = W,Z.
\ee
Here, $M_V$ and $\Gamma_V$ are the gauge-independent and real-valued masses and decay widths identified from the complex pole positions, which differ from their associated counterparts in the real-mass on-shell renormalization scheme \cite{Stuart:1991cc,Freitas:2002ja}. With the introduction of complex gauge-boson masses, the imaginary parts of the charge, wave-function, and mass counterterms must be kept in order to appropriately cancel the UV poles that appear at two-loop order.
We therefore apply on-shell renormalization conditions suitable for the CMS, whose definitions are given in Appendix \ref{appendix_UV}. 
The counterterms related to the input parameters $\alpha$, $\mu_W$, and $\mu_Z$ can also be constructed at ${\cal O}(\alpha\alpha_s)$ from the Feynman integrals presented in~\cite{Dittmaier:2020vra,Kniehl:1989yc,Djouadi:1993ss}.

\subsection{Infrared subtraction}\label{sec:IR}

The IR poles that appear in the UV-renormalized matrix element $\langle \rnM^{(l,m)}|\rnM^{(j,k)}\rangle$ are guaranteed, in a sufficient inclusive observable, to cancel against the corrections containing real gluon and photon emission. Since the soft and collinear IR structure of the amplitude is universal and factorizes from the underlying hard process, it is possible to remove these virtual IR poles via the application of process-independent subtraction operators (see~\cite{Agarwal:2021ais} and references therein).
These subtraction operators fully capture the corresponding structure of the real-emission matrix elements integrated over the phase space of the photon and/or gluon. The final-state collinear divergences are regularized by the mass of the final-state bottom quark, and are proportional to $\log(s/m_b^2)$.

At one-loop order, the finite QCD and EW matrix elements are defined in terms of the IR subtraction operators $\I^{(1,0)}$ and $\I^{(0,1)}$, which act as follows:
\be 
\begin{split}
&\langle\rnM^{(0,0)}|\rnM^{(1,0),fin}\rangle \\
&=\langle\rnM^{(0,0)} |\rnM^{(1,0)}\rangle-\I^{(1,0)}\langle\rnM^{(0,0)}|\rnM^{(0,0)}\rangle,\label{NLO_QCD_finite_amp}
\end{split}
\ee
\be
\begin{split}
&\langle\rnM^{(0,0)}|\rnM^{(0,1),fin}\rangle \\ &=\langle\rnM^{(0,0)}|\rnM^{(0,1)}\rangle-\I^{(0,1)}\langle\rnM^{(0,0)}|\rnM^{(0,0)}\rangle. \label{NLO_EW_finite_amp}
\end{split}
\ee 
Similarly, we define the finite two-loop matrix element as:
\begin{equation}
\begin{split}
&\langle\rnM^{(0,0)}|\rnM^{(1,1),fin}\rangle  \\
&= \langle\rnM^{(0,0)}|\rnM^{(1,1)}\rangle - \I^{(1,1)}\langle\rnM^{(0,0)}|\rnM^{(0,0)}\rangle 
    \\
    &\mathrel{\phantom{=}} -  \I^{(0,1)}\langle\rnM^{(0,0)}|\rnM^{(1,0),fin}\rangle - \I^{(1,0)}\langle\rnM^{(0,0)}|\rnM^{(0,1),fin}\rangle.
    \label{NNLO_finite_amp}
\end{split}
\end{equation}

The two-loop subtraction operator $\I^{(1,1)}$ can be obtained from the NNLO QCD operator\footnote{We use boldface notation for matrices that are non-diagonal in color space.} $\boldsymbol{\I}^{(2,0)}$ for the process $q \bar q \rightarrow Q \bar Q$ \cite{Ferroglia:2009ii} by applying a suitable abelianization procedure \cite{deFlorian:2018wcj,Armadillo:2025mfx}. The result factorizes into the product of the one-loop subtraction operators:
\begin{equation}
\begin{split}
     \I^{(1,1)} = \I^{(1,0)}\I^{(0,1)}.\label{NNLO_IR_sub_factorized}
\end{split}
\end{equation}
The one-loop operators depend on the electric charges $Q_e$ and $Q_b$ and on the Casimir invariant $C_F = \frac{N^2 -1}{2N}$ of the fundamental representation of SU(N). They are given by
\be
 \I^{(1,0)}=\smu^{-\ep}\frac{4}{\ep}\frac{C_F}{2}\l(-1 + \frac{1+x_b^2}{1-x_b^2}\ln (x_b) \r),\label{IR_sub_NLO_QCD}
\ee
\be
\begin{split}
&\I^{(0,1)}=\smu^{-\ep}\bigg\{Q_e^2 \l( - \frac{2}{\ep^2} + \frac{1}{\ep}(3+2i\pi)
\r)\\
&+\frac{4}{\ep}\bigg[Q_e Q_b \ln \l(\frac{2p_1 \cdot p_3}{2p_2 \cdot p_3} \r)+ \frac{Q_b^2}{2}\l(-1 + \frac{1+x_b^2}{1-x_b^2}\ln (x_b) \r) \bigg]\bigg\},\label{IR_sub_NLO_EW}
\end{split}
\ee
where the variable $x_b$ is defined via 
\be
\frac{\l(1-x_b\r 
)^2}{x_b}= - \frac{s}{m_b^2}.
\ee
In contrast to the one-loop QED subtraction operator, the one-loop QCD subtraction operator starts at ${\cal O}(\epsilon^{-1})$ due to the massive nature of the bottom quark. 

The abelianization procedure used to derive Equation \eqref{NNLO_IR_sub_factorized} can be summarized as follows. First, terms in $\boldsymbol{\I}^{(2,0)}$ associated with one-loop corrections to the emitted gluon are dropped, since corrections of this type are impossible at mixed order (e.g., there are no one-loop EW corrections to the gluon propagator or one-loop QCD corrections to the photon propagator). Secondly, pieces built from products of the one-loop anomalous-dimension matrix $\boldsymbol{\mathrm{\Gamma}}$ and its derivative $\mathrm{\Gamma}'$
are mapped onto \oaas via
\be 
\boldsymbol{\mathrm{\Gamma}}^{(1,0)}\mathrm{\Gamma}'^{(1,0)} \rightarrow \boldsymbol{\mathrm{\Gamma}}^{(1,0)}\mathrm{\Gamma}'^{(0,1)}+\mathrm{\Gamma}^{(0,1)}\mathrm{\Gamma} '^{(1,0)},
\ee
with analogous replacements for $\boldsymbol{\mathrm{\Gamma}}^{(1,0)}\boldsymbol{\mathrm{\Gamma}}^{(1,0)}$ and $\mathrm{\Gamma}'^{(1,0)}\mathrm{\Gamma}'^{(1,0)}$. Lastly, genuine two-loop terms are projected onto their mixed-coupling counterparts through
\be
C_F^2 \rightarrow 2 Q_q^2 C_F, \ \ \ \ C_F C_A \rightarrow 0, \ \ \ \ C_F T_F \rightarrow 0,
\ee 
where the factor of $2$ is necessary since there are two possible orderings of simultaneous photon and gluon exchange that contribute to the quark self-energy at $\mathcal O(\alpha \alpha_s)$.
By inspection, we observe that the IR structure of the contributions containing both one-loop initial-state and one-loop final-state vertex corrections factorizes. The same is true of the two-loop box diagrams, since gluons and photons do not interact with each other, and the initial-state electrons are colorless. In the case of the two-loop final-state vertex corrections, the bottom-quark mass regularizes the collinear singularities, so that only soft divergences are possible, and these factorize into the product of two single-exchange poles.

\section{The amplitude at \texorpdfstring{${\cal O}(\alpha^2 \alpha_s)$}{O(alpha2 alphaS)}\label{amp}}

\begin{figure}[t]
    \centering
    \begin{subfigure}[c]{0.15\textwidth}
        \centering
        \includegraphics[width=\linewidth]{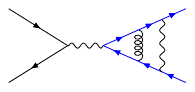}
        \caption{}\label{fig1a}
    \end{subfigure}
    \begin{subfigure}[c]{0.15\textwidth}
        \centering
        \includegraphics[width=\linewidth]{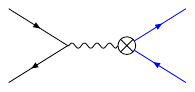}
        \caption{}\label{fig1b}
    \end{subfigure}
    \begin{subfigure}[c]{0.15\textwidth}
        \centering
        \includegraphics[width=\linewidth]{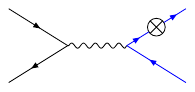}
        \caption{}\label{fig1c}
    \end{subfigure}

    \medskip

    \begin{subfigure}[c]{0.15\textwidth}
        \centering
        \includegraphics[width=\linewidth]{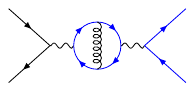}
        \caption{}\label{fig1d}
    \end{subfigure}
    \begin{subfigure}[c]{0.15\textwidth}
        \centering
        \includegraphics[width=\linewidth]{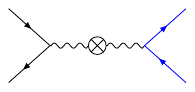}
        \caption{}\label{fig1e}
    \end{subfigure}
    \begin{subfigure}[c]{0.15\textwidth}
        \centering
        \includegraphics[width=\linewidth]{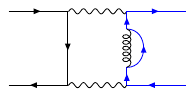}
        \caption{}\label{fig1f}
    \end{subfigure}
    \caption{Sample Feynman diagrams\footnote{Arrowed lines represent fermions, wavy lines represent EW gauge bosons, and coiled lines represent gluons. Black lines represent massless particles, while different colored lines represent particles of different mass. Note that any photon line can be replaced with a $Z$-boson line to obtain an additional contributing diagram.} of two-loop corrections and associated two-loop counterterms.}
    \label{fig1}
\end{figure}

\begin{figure}
   \centering
    \begin{subfigure}[c]{0.15\textwidth}
        \centering
        \includegraphics[width=\linewidth]{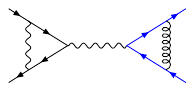}
        \caption{}\label{fig2a}
    \end{subfigure}
    \begin{subfigure}[c]{0.15\textwidth}
        \centering
        \includegraphics[width=\linewidth]{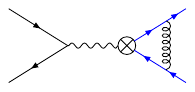}
        \caption{}\label{fig2b}
    \end{subfigure}
    \begin{subfigure}[c]{0.15\textwidth}
        \centering
        \includegraphics[width=\linewidth]{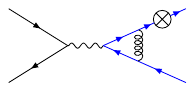}
        \caption{}\label{fig2c}
    \end{subfigure}

    \medskip

    \begin{subfigure}[c]{0.15\textwidth}
        \centering
        \includegraphics[width=\linewidth]{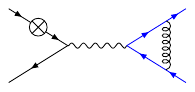}
        \caption{}\label{fig2d}
    \end{subfigure}
    \begin{subfigure}[c]{0.15\textwidth}
        \centering
        \includegraphics[width=\linewidth]{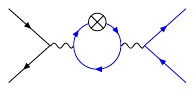}
        \caption{}\label{fig2e}
    \end{subfigure}
    \begin{subfigure}[c]{0.15\textwidth}
        \centering
        \includegraphics[width=\linewidth]{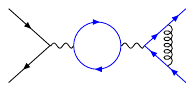}
        \caption{}\label{fig2f}
    \end{subfigure}
    
    \medskip

    \begin{subfigure}[c]{0.15\textwidth}
        \centering
        \includegraphics[width=\linewidth]{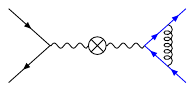}
        \caption{}\label{fig2g}
    \end{subfigure}
    \begin{subfigure}[c]{0.15\textwidth}
        \centering
        \includegraphics[width=\linewidth]{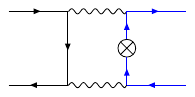}
        \caption{}\label{fig2h}
    \end{subfigure}       
        \caption{Sample Feynman diagrams of factorisable and one-loop counterterm corrections.}
        \label{fig2}
\end{figure}

A large number of Feynman diagrams contribute to the scattering amplitude at ${\cal O}(\alpha^2 \alpha_s)$. We provide few representative examples
in Figures~\ref{fig1} and \ref{fig2}.
\begin{figure}
   \centering
    \begin{subfigure}[c]{0.15\textwidth}
        \centering
        \includegraphics[width=\linewidth]{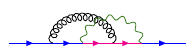}
        \caption{}\label{fig3a}
    \end{subfigure}
    \begin{subfigure}[c]{0.15\textwidth}
        \centering
        \includegraphics[width=\linewidth]{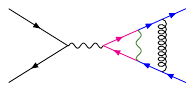}
        \caption{}\label{fig3b}
    \end{subfigure}
    \begin{subfigure}[c]{0.15\textwidth}
        \centering
        \includegraphics[width=\linewidth]{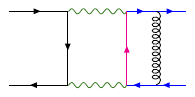}
        \caption{}\label{fig3c}
    \end{subfigure}

        \caption{Sample Feynman diagrams of two-loop corrections that contain three different massive propagators represented by different colors.}
        \label{fig3}
        \centering
    \end{figure}
The mixed QCD-EW corrections to $e^+e^-\to b\bar{b}$ are characterized by the fact that the virtual gluon has to be exchanged along the final state bottom-quark line, whereas the EW bosons can be attached to the initial, final, and between the initial and final states in all possible ways.
At the level of 1PI corrections,
we find non-vanishing exchanges of scalar particles only in the final-state vertex diagrams, because of the massive nature of the bottom quark; the initial-state vertex and all box corrections feature only the exchange of vector bosons.
The two-loop gauge-boson self-energies are characterized by the presence of a closed quark loop with the internal exchange of a gluon. We do not find two-loop mixed QCD-EW wave-function corrections to the incoming electron line, because of its color-neutral nature, while they are present in the final state.
Considering the factorizable contributions with two one-loop corrections (either closed loop or counterterm insertion), one of the two has to be proportional to the strong coupling and has to be along the bottom-quark line.

The complexity of the Feynman integrals present in the amplitude can be classified according to the number of internal massive lines and the number of distinct masses present in the same integral. Notably, the bottom self-energy corrections (which contribute to the external-leg renormalization), the vertex corrections, and the box corrections feature integrals that contain 4 or 5 massive lines with 3 distinct masses. Sample corrections of this type are given in Figure~\ref{fig3}.

All scalar integrals that appear in the calculation are characterized by the integral family to which they belong. An integral family is defined by the specific combinations of external and loop momenta that appear in the propagator denominators, as well as by which of these denominators contain masses.
A total of 35 distinct integral families are identified: 17 arise from box diagrams, 13 from vertex-correction diagrams, 1 from self-energy diagrams, and 4 from factorizable diagrams.
We provide the details of each integral family in Appendix~\ref{appendix_families}.

After we obtain the necessary IBP reduction rules as described in Section~\ref{sec:metho} and apply them, we find a total of 1621 master integrals. 
The two-loop box/vertex/factorizable/self-energy integral families contribute 913/656/50/2 of the total number of master integrals, respectively.
As a result, we find the bare amplitude can be reduced to a linear combination of MIs $I_k$ and corresponding rational coefficients $c_k$:
\begin{equation}
\begin{split}
&\langle\rnM^{(0,0)}_0|\rnM^{(1,1)}_0\rangle\\
&=\sum_{k=1}^{1621}c_k(s,t,\{m_f\},\{\mu_B\},\epsilon)I_k(s,t,\{m_f\},\{\mu_B\},\epsilon),\label{2L_sum_ck_Ik}
\end{split}
\end{equation}
where  $\{m_f\}$,  $\{\mu_B\}$ are the sets of the masses of the fermions and bosons, respectively. 
Because the coefficients $c_k$ are a rational function of their arguments, they can be evaluated directly to high precision. 
As mentioned in Section~\ref{amp}, we use \AMFlow to evaluate the MIs. Computing all two-loop master integrals at a single kinematic point takes approximately six days on the PCEPC cluster, the information of which can be found in Appendix \ref{appendix_cluster}. 
An effective approach to compute the MIs  on a large-scale grid of kinematic points is to solve the system of differential equations satisfied by them with the series expansion method~\cite{Moriello:2019yhu,Hidding:2020ytt,Liu:2022chg,Huang:2026rjb}.
To prepare the necessary tools for large-scale grid scanning, we obtained the differential equations in the Mandelstam variables for each integral family.
The size of the differential equations can reach GB level for the most complicated families.

\section{Cross-checks}
\label{cross-checks}
\subsection{Cancellation of the poles}\label{sec:subtr}

The main result of this paper is our computation of the UV-renormalised and IR-subtracted matrix element $\langle \rnM^{(0,0)}| \rnM^{(1,1),fin}\rangle$ at a benchmark kinematic point.
In this section, we describe the verification of the UV- and IR-pole cancellation we have obtained while extracting the finite contribution.
The possibility of performing such a check with an arbitrary number of significant digits provides strong evidence for the validity of our computational framework. 

To proceed, let $Q$ be the order of magnitude of the largest numerical term $X$ appearing in the calculation, $Q=\log_{10}(X)$.
We then consider the cancellation of the poles successful at a given order if, for any value of digits requested by our arbitrary numerical precision $P$, the sum of the numerical coefficients is less than or equal to our numerical representation of zero, $10^{-(P-Q)}$.
In our calculation, we fix $P=70$. However, cancellations at intermediate stages result in a loss of precision: we find $Q=14$. 
Therefore, our numerical representation of zero is $10^{-56}$. 

With this criterion, we investigated the cancellation of the poles in the box, vertex-correction, factorizable, and self-energy subsets of the ${\cal O}(\alpha^2 \alpha_s)$ virtual corrections individually.
Consider first the two-loop bare and one-loop counterterm contributions to the gauge-boson self-energies (Figures \ref{fig1d} and \ref{fig2e}), which are IR finite. We have confirmed that the UV poles in the sum of these contributions cancel with the insertion of two-loop mass counterterms (Figure \ref{fig1e}) and two-loop charge counterterms at the adjacent vertices (Figure \ref{fig1b}), as expected in the BFM due to the Ward identities (see Equations \eqref{AA_WF_CT}-\eqref{AZ_WF_CT}).

The box, vertex-correction, and factorizable subsets, on the other hand,
are both UV and IR divergent. At two-loop order, the UV and IR poles become entangled, and their cancellation is non-trivial. Nevertheless,
we can identify the IR structure of each subset by observing the distinct combinations of couplings that appear in the subtraction terms of Equation \eqref{NNLO_finite_amp}. For instance, the IR structure of the final-state vertex corrections is given by the terms proportional to $C_F Q_b^2$. We have confirmed that combining these terms with the sum of the two-loop bare and one-loop counterterm vertex corrections (Figures \ref{fig1a} and \ref{fig2c}) and the two-loop external quark wave-function corrections (Figure \ref{fig1c}) yields a finite result. The UV finiteness of this sum is also guaranteed by the BFM Ward identities.  For a numerical demonstration of the IR pole cancellation in the full set of corrections, we refer the reader to Table \ref{benchmark_poles}.

\subsection{Treatment of \texorpdfstring{$\gamma_5$}{gamma5}}\label{sec:gamma5}
As described in Section~\ref{sec:metho}, the independence of our results from the ambiguities inherent in the adopted prescription for treating $\gamma_5$ in dimensional regularization provides additional validation of our computational framework.

In this calculation, we follow Kreimer's prescription and preserve the anticommuting behavior of $\gamma_5$ in $d$ dimensions, losing in turn the trace cyclicity. This choice strongly reduces the computational load, because the trace of each fermion line contains zero or at most one $\gamma_5$ matrix. The latter is then replaced by its covariant 4-dimensional representation, 
\begin{equation}
    \gamma_5 = \frac{i}{4!} \epsilon^{\mu\nu\rho\sigma} \gamma_\mu \gamma_\nu \gamma_\rho \gamma_\sigma,
\end{equation}
and the resultant traces can be computed in $d$ dimensions without issue, since they contain no $\gamma_5$. The Levi-Civita tensor is defined to be a 4-dimensional object, avoiding any ambiguity when two such tensors are contracted. 

In Kreimer's prescription, the non-cyclicity of the trace requires the adoption of conventions that specify the ordering of the internal arguments. The ``reading-point" convention determines the point from which internal fermion traces start, and the ``fixed-point" convention dictates how the spinors are ordered in the traces that arise from the external polarization sums. Different choices of the conventions may lead to different intermediate results. However, the UV-renormalized and IR-subtracted matrix element is independent of these choices once all contributions are combined.

 \abiss allows the user to select their preferred definition for each ordering convention.
At ${\cal O}(\alpha^2 \alpha_s)$, all corrections featuring fermion triangles vanish due to color algebra, so fermionic traces that are sensitive to the convention for $\gamma_5$ only arise from the sums over external polarization states.

Consider an arbitrary interference term characterized by the matrices $A$, $B$, $C$, and $D$. We define the only two independent external spinor orderings as: 
\begin{enumerate} 
\item[I)] \hypertarget{ordering_I}{Ordering $\{ p_1, p_2, p_3, p_4 \}$}:
    \begin{align}
    \sum_{r_1,r_2,s_1,s_2} &\left[\left(\bar{u}_{r_1}(p_1) A v_{r_2}(p_2)\right)\left(\bar{v}_{r_2}(p_2) B u_{r_1}(p_1)\right)\right] \nonumber \\
    &\left[\left(\bar{u}_{s_1}(p_3) C  v_{s_2}(p_4)\right)\left(\bar{v}_{s_2}(p_4) D u_{s_1}(p_3)\right)\right]
     \nonumber \\
   = \ & \text{Tr}\left[(\slashed{p}_1 + m_1)A (\slashed{p}_2 - m_2) B \right]  \nonumber \\
   & \text{Tr}\left[ (\slashed{p}_3 + m_3) C (\slashed{p}_4 - m_4) D \right] \nonumber
    \end{align} 
\item[II)] \hypertarget{ordering_II}{Ordering $\{ p_2, p_1, p_3, p_4 \}$}:
    \begin{align}
    \sum_{r_1,r_2,s_1,s_2} &\left[\left(\bar{v}_{r_2}(p_2) B u_{r_1}(p_1)\right)\left(\bar{u}_{r_1}(p_1) A v_{r_2}(p_2)\right)\right] \nonumber \\
    &\left[\left(\bar{u}_{s_1}(p_3) C  v_{s_2}(p_4)\right)\left(\bar{v}_{s_2}(p_4) D u_{s_1}(p_3)\right)\right]
     \nonumber \\
    = \ & \text{Tr}\left[(\slashed{p}_2 - m_2) B (\slashed{p}_1 + m_1)A \right] \nonumber \\
    & \text{Tr}\left[ (\slashed{p}_3 + m_3) C (\slashed{p}_4 - m_4)D\right] \nonumber 
    \end{align} 
\end{enumerate}
As a numerical and methodological test, we computed the UV-renormalized and IR-subtracted matrix element under both orderings ``I" and ``II". We found the two-loop vertex and self-energy corrections to coincide in the two cases, and the difference corresponding to the two-loop box corrections, $\langle \mathcal{M}^{(0,0)}| \mathcal{M}^{(1,1),fin}_{\text{box}}\rangle$, to vanish in the limit $\epsilon \rightarrow 0$:
\begin{align}
    & \langle \mathcal{M}^{(0,0)}| \mathcal{M}^{(1,1),fin}_{\text{box}}\rangle_\text{I} - \langle \mathcal{M}^{(0,0)}| \mathcal{M}^{(1,1),fin}_{\text{box}}\rangle_\text{II} \nonumber \\
    & = \left( \langle \mathcal{M}^{(0,0)}| \mathcal{M}^{(1,1)}_{\text{box}}\rangle_\text{I} - \langle \mathcal{M}^{(0,0)}| \mathcal{M}^{(1,1)}_{\text{box}}\rangle_\text{II} \right) 
    \nonumber \\
    &\mathrel{\phantom{=}} - \I^{(1,1)}_{C_F Q_e Q_b}\left(\langle\rnM^{(0,0)}|\rnM^{(0,0)} \rangle_\text{I}- \langle\rnM^{(0,0)}|\rnM^{(0,0)} \rangle_\text{II}\right) \nonumber \\ 
    &\mathrel{\phantom{=}} - \I^{(1,0)}\left(\langle\mathcal{M}^{(0,0)}|\mathcal{M}^{(0,1),fin}_{\text{box}}\rangle_\text{I} - \langle \mathcal{M}^{(0,0)}| \mathcal{M}^{(0,1),fin}_{\text{box}}\rangle_\text{II} \right) \nonumber \\
    &\mathrel{\phantom{=}}- \tilde\I^{(0,1)}_{Q_e Q_b}\left(\langle \rnM^{(0,0)}|\rnM^{(1,0),fin}\rangle_\text{I}-\langle \rnM^{(0,0)}|\rnM^{(1,0),fin}\rangle_\text{II} \right)\nonumber \\
    & = \mathcal{O}(\epsilon),  \label{eq:2L_diff_fixedpoint}
\end{align} 
which provides a strong check of the internal consistency of our treatment of chiral couplings. We therefore claim that our results are independent of any possible ambiguity following from the adopted prescription for $\gamma_5$. For a numerical demonstration of the result in Equation \eqref{eq:2L_diff_fixedpoint}, we refer the reader to Table \ref{benchmark_gamma5}.

\section{Benchmark results} \label{results}

\begin{table*}
\renewcommand{\arraystretch}{1.3}
\begin{tabular}{ c|c|c } 
\Xhline{1pt}
 Powers of $\epsilon$  & Individual coefficients & Net coefficient \\ 
\hline
    & $-96.0374448313433847541171432268+61.2866236749924468940553952081~i$   &\\ \cline{2-2}
$\epsilon^{-3}$&  $96.0374448313433847541171432268-61.2866236749924468940553952081~i$ & $10^{-59}+ 10^{-59}~i$ \\ \cline{2-2}
   & $0$ & \\ 
   \hline
   & $549.64698096126642874719163466-1042.90986679350434825886500505~i$ &\\ \cline{2-2}
$\epsilon^{-2}$ & $-1039.99871537577438129779093702+1135.43392566315541196550978211~i$ & $10^{-57}+10^{-58}~i$  \\ \cline{2-2}
   & $490.35173441450795255059930236-92.52405886965106370664477706~i$ & \\ \hline
   & $-7368.3706389329137333807213566+5351.6460679711039989141578487~i$ & \\
   & $+(603.4210623033668745288841917-385.0752134013571344891444095~i)\Delta\alpha_\text{had}^{(5)}(M^2_{Z,\text{OS}})/\alpha$ & \\ \cline{2-2}
$\epsilon^{-1}$ & $4615.6738545989896322390967459-10104.2147016345237628632516080~i $ & $10^{-56}+10^{-56}~i $  \\ \cline{2-2}
  & $2752.6967843339241011416246106+4752.5686336634197639490937592~i$ & \\
  & $-(603.4210623033668745288841917-385.0752134013571344891444095~i)\Delta\alpha_\text{had}^{(5)}(M^2_{Z,\text{OS}})/\alpha$ & \\
\Xhline{1pt}
\end{tabular}
\caption{Coefficients of the 
$\epsilon^{-k}$ poles appearing in each contribution to Equation \eqref{NNLO_finite_amp} at the benchmark kinematic point $s = 6561 \ \text{GeV}^2$, $t = - 956 \ \text{GeV}^2$ expressed in units of $N_C\alpha^3 \alpha_s$ with at most 30 significant digits to maintain readability. For each $k$, the first row corresponds to $\langle \rnM^{(0,0)} | \rnM^{(1,1)}\rangle$, the second to $-\I^{(1,1)}\langle \rnM^{(0,0)}|\rnM^{(0,0)}\rangle$, and the third to
$- \I^{(0,1)}\langle \rnM^{(0,0)}| \rnM^{(1,0),fin}\rangle-\I^{(1,0)}\langle \rnM^{(0,0)}| \rnM^{(0,1),fin}\rangle$. In the final column we present the sum of the poles, numerically consistent with zero given the available precision.}
\label{benchmark_poles}
\end{table*}

In this section we present benchmark results of our calculation, evaluated at a specific kinematic point, in order to demonstrate the validity of the checks performed on the amplitude.
The following values for the input parameters were used in the numerical evaluation of the MIs and their coefficients:
\begin{center}
\renewcommand{\arraystretch}{1.3}
\begin{tabular}{l@{~~~~~~}l}
\hline\hline
$m_e = 0.510998 \ \text{MeV}$ & $M_Z = 91.1535 \ \text{GeV}$ \\
$m_\mu = 0.105660 \ \text{GeV}$ & $\Gamma_Z = 2.49430 \ \text{GeV}$ \\
$m_\tau = 1.77686 \ \text{GeV}$ & $M_W = 80.3580\ \text{GeV}$ \\
$m_b = 4.18000 \ \text{GeV}$ & $\Gamma_W = 2.08400 \ \text{GeV}$ \\
$m_t = 172.690 \ \text{GeV}$ & $M_{Z,\text{OS}} = 91.1880\ \text{GeV}$\\
\hline\hline
\end{tabular}
\end{center}
We treat the initial-state electrons as massless.
In the gauge-boson self-energies, all leptons are kept massive, but the masses of the four lightest quarks are explicitly set to zero. In addition, we replace the CKM matrix with the identity matrix. 
Numerical results at the benchmark kinematic point $s = 6561 \ \text{GeV}^2$, $t = - 956 \ \text{GeV}^2$,
chosen away from the $Z$ resonance, are provided in Table \ref{benchmark_poles}, Equations \eqref{fin_res}-\eqref{H11_res}, and Table \ref{benchmark_gamma5},
expressed in units of $N_C\alpha^3 \alpha_s$. To illustrate the cancellation of the IR poles, Table \ref{benchmark_poles} lists the coefficients of $\epsilon^{-k}$ appearing in the Laurent expansion of each contribution to $\langle\rnM^{(0,0)}|\rnM^{(1,1),fin}\rangle$ in Equation \eqref{NNLO_finite_amp}. As shown in the final column, the coefficients of each pole cancel to 
$10^{-56}$ or smaller, satisfying the criterion discussed in Section \ref{sec:subtr}. 
On the other hand,
summing the finite parts of each term in Equation \eqref{NNLO_finite_amp} yields the benchmark value for the UV-renormalized and IR-subtracted matrix element:
\begin{align}
&\langle\rnM^{(0,0)}|\rnM^{(1,1),fin}\rangle\nonumber\\
&= -21678.1718967267798126094003239 \nonumber \\
&\mathrel{\phantom{=}}-21293.8829011884802740244402874~i\nonumber \\
&\mathrel{\phantom{=}}+(3080.9708130232631555382840879\nonumber\\
&\mathrel{\phantom{=}}-581.3458072504106614617398866~i)\Delta\alpha_\text{had}^{(5)}(M^2_{Z,\text{OS}})/\alpha.\label{fin_res}
\end{align}
This corresponds to the hard function contribution
\begin{align}
   & H^{(1,1)}\nonumber\\
   & =-740.835523997346285441232582213\nonumber \\
&\mathrel{\phantom{=}}+105.289903482648193435962778129\Delta\alpha_\text{had}^{(5)}(M^2_{Z,\text{OS}})/\alpha.\label{H11_res}
\end{align}
The presence of $\Delta\alpha_\text{had}^{(5)}(M_{Z,\text{OS}}^2)$ in Table \ref{benchmark_poles} and Equations \eqref{fin_res} and \eqref{H11_res} is due to the non-perturbative nature of the light-quark contribution to the on-shell electric charge counterterm (see Equation \eqref{pi_AA_split}). We keep the dependence on $\Delta\alpha_\text{had}^{(5)}(M_{Z,\text{OS}}^2)$ parametric to ease the evaluation of its impact in future phenomenological studies.

Furthermore, to demonstrate the independence of our results from the ambiguities present in $\gamma_5$ Kreimer's prescription, we compare the values obtained with the two different spinor orderings described in Section \ref{sec:gamma5} for the non-trivial case of the box diagram contributions. Table \ref{benchmark_gamma5} lists the coefficients of $\epsilon^{-k}$ and $\epsilon^0$ in the Laurent expansion of each contribution to $  \langle \mathcal{M}^{(0,0)}| \mathcal{M}^{(1,1),fin}_{\text{box}}\rangle$ 
in Equation \eqref{eq:2L_diff_fixedpoint}. The finite part of the UV-renormalized but IR-divergent matrix element $\langle \mathcal{M}^{(0,0)}| \mathcal{M}^{(1,1)}_{\text{box}
}\rangle$ differs between spinor orderings \hyperlink{ordering_I}{I} and \hyperlink{ordering_II}{II}. However, the one-loop finite remainder $\langle \rnM^{(0,0)}| \rnM^{(0,1),fin}_{\text{box}}\rangle$ only differs starting at ${\cal O}(\epsilon)$. When multiplied by the one-loop QCD subtraction operator, 
this difference exactly compensates that in $\langle \mathcal{M}^{(0,0)}| \mathcal{M}^{(1,1)}_{\text{box}
}\rangle$, canceling all dependence on the spinor ordering.

\begin{table*}
\renewcommand{\arraystretch}{1.3}
\begin{tabular}{ c|c|c } 
\Xhline{1pt}
 Powers of $\epsilon$  & Spinor ordering I  & Spinor ordering II \\ 
\hline
    &$-111.87867444499615+71.39575849824868~i$  & $-111.87867444499615+71.39575849824868~i$\\ \cline{2-3}
$\epsilon^{-2}$ &  $111.87867444499615-71.39575849824868~i$ & $111.87867444499615-71.39575849824868~i$ \\ \cline{2-3}
   & $0$ & $0$\\ 
   \hline
   & $414.2026774143816-521.8700903387290~i$ & $414.2026774143816-521.8700903387290~i$\\ \cline{2-3}
$\epsilon^{-1}$ & $-1697.7165848553345+1083.4036414186723~i$ & $-1697.7165848553345+1083.4036414186723~i$  \\ \cline{2-3}
   & $1283.5139074409529-561.5335510799433~i$ & $1283.5139074409529-561.5335510799433~i$\\ \hline
   & $-1082.578129337419+1953.168032413456~i$ & $-912.016346039799+2026.456264089261~i$ \\ \cline{2-3}
$\epsilon^{0}$ & $12754.493508174951-8139.323627085281~i$ & $12754.493508174951-8139.323627085281~i$  \\ \cline{2-3}
  & $-15232.585522220781+6825.157089559418~i$ & $-15403.147305518401+6751.868857883613~i$\\
  \hline
  $\langle \rnM^{(0,0)} | \rnM^{(1,1),fin}_{\text{box}}\rangle$ & $-3560.670143383249+639.001494887593~i$ & $-3560.670143383249+639.001494887593~i$\\
\Xhline{1pt}
\end{tabular}
\caption{Coefficients of 
$\epsilon^{-k}$ and $\epsilon^0$ appearing in each contribution to Equation \eqref{eq:2L_diff_fixedpoint} at the benchmark kinematic point $s = 6561 \ \text{GeV}^2$, $t = - 956 \ \text{GeV}^2$ expressed in units of $N_C\alpha^3 \alpha_s$ with at most 17 significant digits to maintain readability. For each $k$, the first row corresponds to $\langle \rnM^{(0,0)} | \rnM^{(1,1)}_{\text{box}}\rangle$, the second to $-\I^{(1,1)}_{C_FQ_eQ_b}\langle\rnM^{(0,0)}|\rnM^{(0,0)}\rangle$, and the third to
$-  \I^{(0,1)}_{Q_eQ_b}\langle \rnM^{(0,0)}| \rnM^{(1,0),fin}\rangle-\I^{(1,0)}\langle \rnM^{(0,0)}| \rnM^{(0,1),fin}_{\text{box}}\rangle$, computed with spinor orderings \protect\hyperlink{ordering_I}{I} and \protect\hyperlink{ordering_II}{II}. When comparing the two spinor orderings, the sum of the coefficients in the three rows is the same for each power of $\epsilon$.}
\label{benchmark_gamma5}
\end{table*}

\twocolumngrid
\section{Conclusions} \label{conclusions}
In this work we have presented the evaluation of the $\mathcal{O}(\alpha^2\alpha_s)$ contributions to the scattering amplitude of the process
$e^+ e^-\rightarrow b\bar{b}$ with massive bottom quarks.
The results are obtained following an automated workflow, described in Section \ref{sec:metho}, which allows a consistent calculation of the unpolarized matrix elements of the process; the latter are eventually cast as a linear combination of master integrals, each multiplied by a rational function coefficient. The numerical evaluation relies on standard tools, and features results with a precision arbitrarily set by the user.
This high precision offers in turn a very high level of control over every single step of the calculation.
The final output has been extensively tested and satisfies several necessary conditions: the cancellation of UV and IR poles, the independence from the conventions adopted to handle the traces involving a $\gamma_5$ matrix, and more formal checks like the validity of the Ward identities for the Green's functions which contribute to the amplitude.

All of these properties have been illustrated at one kinematic point.
The preparation of a numerical grid allows to efficiently perform
phenomenological studies:
the correction at an arbitrary phase-space point can be obtained via interpolation of the grid values, with an accuracy that can be guaranteed at will, including a sufficiently fine-grained scan in the grid preparation.
Such grids are conveniently evaluated by solving the differential equations satisfied by each integral family with \seasyde \cite{Armadillo:2022ugh}.
A phenomenological study is in preparation and left to a forthcoming publication.

The methodology presented in this paper allows several future applications, like the evaluation of mixed QCD-EW corrections to the production of a top-antitop pair, or the complete set of two-loop virtual EW corrections to fermion pair production in electron-positron annihilation.

\vskip 0.2in
 
\noindent {\bf{Acknowledgments}}
We thank Z. Mo for useful discussions during this work. C. Dey, Z. Huang, X. Li, M. J. Ramsey-Musolf, A. Vicini, and J. Zhu were supported in part under National Natural Science Foundation of China (NSFC) grant W2441004. M. J. Ramsey-Musolf and J. Zhu were also supported in part under NSFC grant 12375094. 
A. Vicini is supported by the European Research Council (ERC) under grant agreements 101267906 (ERC Advanced Grant Quantum FronTiers).
The work of S. Devoto received support from the FWO (contract No. 1227426N).
A. Vicini is grateful to the TDLI Institute of the Shanghai Jiao Tong University for the support and hospitality during a 4-month stay when this work was prepared.
J. T. Fagnoni is grateful to INFN Milan Unit for the support and hospitality during a 2-month stay when this work was prepared.
\appendix

\section{UV counterterms}\label{appendix_UV}

In this appendix, we provide the definitions for the relevant counterterms in the on-shell complex-mass scheme. In the gauge sector, the unrenormalized self-energies are decomposed as
\be
\begin{split}
\Sigma^{VV'}_{\mu\nu}(k) &= -\left(  g_{\mu\nu} - \frac{k_\mu k_\nu}{k^2}\right) \Sigma^{VV'}_T(k^2)\\
& \phantom{=}\  - \frac{k_\mu k_\nu}{k^2}\Sigma_L^{VV'}(k^2).
\end{split}
\ee
The mass counterterms $\delta\mu_W^2$ and $\delta \mu_Z^2$ are obtained from the transverse parts evaluated at the corresponding complex poles: 
\be
\delta \mu_W^2 = \Sigma_T^{WW}(\mu_W^2), \ \ \ \ \delta\mu_Z^2 = \Sigma_T^{ZZ}(\mu_Z^2).
\ee
The wave-function counterterms are also defined in terms of the transverse self-energies, but, in the background-field formalism, they are linked directly to the charge 
counterterms 
through ${\cal O}(\alpha\alpha_s)$:
\begin{align}
    \delta Z_{AA} &= - 2\frac{\delta e}{e},\label{AA_WF_CT}\\
    \delta Z_{ZZ} &= - 2 \frac{\delta e}{e}- \frac{s_W^2 - c_W^2}{c_W^2}\frac{\delta s_W^2}{s_W^2},\label{ZZ_WF_CT}\displaybreak[0]\\
    \delta Z_{AZ}&=-2 \frac{\delta s_W^2}{c_W s_W},\label{AZ_WF_CT}\\
    \delta Z_{ZA}&=0,
\end{align}
where $c_W^2 = 1- s_W^2 =\mu_W^2/\mu_Z^2$. 
However, we do not introduce wave-function counterterms for internal fields. Instead, Equations \eqref{AA_WF_CT}-\eqref{AZ_WF_CT} imply that the combination of gauge-boson self-energies and mass counterterms is rendered UV finite by the insertion of charge 
counterterms at the vertices on either side of the corresponding propagators.

Making use of Equation \eqref{AA_WF_CT}, $\delta e$ is obtained directly from the photon vacuum polarization 
\be
\Pi^{AA}(0) =\left.\frac{\partial \Sigma^{AA}_T(k^2)}{\partial k^2}\right|_{k^2 = 0},
\ee
which is sensitive to non-perturbative QCD. To this end, we separate $\Pi^{AA}(0)$ as 
\begin{equation}
\begin{split}
    \mathrm{\Pi}^{AA}(0) &= \mathrm{\Pi}^{AA}_\text{lep}(0)+\mathrm{\Pi}^{AA}_\text{bos}(0) + \mathrm{\Pi}^{AA(t)}_\text{had}(0) \\
    &+ \text{Re}\left[ \mathrm{\Pi}^{AA(5)}_{\text{had}}(M_{Z,\text{OS}}^2)\right] + \mathrm{\Delta} \alpha_\text{had}^{(5)}(M_{Z,\text{OS}}^2)\label{pi_AA_split},
\end{split}
\end{equation}
where 
$M_{Z,\text{OS}}$
is the on-shell mass of the $Z$ boson. The first four terms in Equation \eqref{pi_AA_split} are evaluated perturbatively, taking the four lightest quarks to be massless. 
The final term
is finite and captures all of the non-perturbative QCD input. Its value, which contains effects beyond any fixed-order calculation, can be extracted from experimental data. For correct perturbative bookkeeping, however, it must be treated as an ${\cal O}(\alpha)$ contribution to $\delta e$.

Because the initial-state electrons are taken to be massless, the two-loop counterterms in the scalar sector are not needed to renormalize the amplitude. In addition, the insertion of one-loop scalar counterterms into one-loop diagrams would yield ${\cal O}(\alpha^3)$ corrections rather than ${\cal O}(\alpha^2 \alpha_s)$. We therefore omit their definitions here, which  leaves only the fermion counterterms. The unrenormalized fermion self-energies are decomposed as 
\be
\begin{split}
\Sigma^{f}(k) = \slashed k\left[\omega_-\Sigma^{f}_L(k^2)+\omega_+ \Sigma^{ f}_R(k^2)\right]+ m_f \Sigma^f_S(k^2),
\end{split}
\ee
where $\omega_\pm = (1\pm \gamma_5)/2$ are the chiral projectors. The mass and wave-function counterterms are obtained from the self-energies as follows:
\begin{align}
&\delta m_f = \frac{m_f}{2} \left[\Sigma^f_L(m_f^2) + \Sigma^f_R(m_f^2) + 2 \Sigma^f_S(m_f^2) \right], \nonumber\\
&\delta Z^f_{L/R} = - \Sigma^f_{L/R}(m_f^2)\nonumber \\
&\phantom{=}\ -m_f^2 \frac{\partial }{\partial k^2}\left.\left[ \Sigma^f_L(k^2) + \Sigma^f_R(k^2) + 2 \Sigma^f_S(k^2) \right]\right|_{k^2 = m_f^2}.
\end{align}
The fermion wave-function counterterms are necessary to renormalize the external legs of the initial and final states.

Taking the on-shell limit in the definitions of the electric charge and fermion counterterms requires additional care. To illustrate this, let $f(k^2)$ represent an arbitrary self-energy function or its derivative, decomposed into coefficients $c_j$ and MIs $I_j$ as
\be
f(k^2) = \sum_j c_j(k^2) I_j(k^2),
\ee
and let $k^2 = M^2$ represent the corresponding on-shell condition. Even though $f(M^2)$ is well-defined, the coefficients $c_j$ can contain factors of $(k^2 - M^2)^{-1}$. The MIs $I_j$ must therefore be expanded in powers of $(k^2- M^2)$ before taking the on-shell limit:
\be
I_j(k^2) \rightarrow \sum_{n=0}^{n_\text{max}} \frac{1}{n!}\left.\frac{\partial^n I_j(k^2)}{\partial(k^2)^n}\right|_{k^2=M^2}(k^2-M^2)^n,
\label{MI_expansion}
\ee
where $n_\text{max}$ is the highest power of $(k^2-M^2)^{-1}$ that appears in $f(k^2)$. We use the \mathematica package \litered to perform the expansion in Equation \eqref{MI_expansion} with $n_\text{max}=2$, which is sufficient at ${\cal O}(\alpha \alpha_s)$.

\section{Summary of Integral Families}\label{appendix_families}
\begin{figure}[t]
 \renewcommand{\thesubfigure}{\Alph{subfigure}}
    \centering
    \begin{subfigure}[c]{0.15\textwidth}
        \centering
        \includegraphics[width=\linewidth]{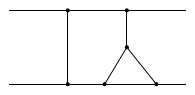}
        \caption{}\label{fig5a}
    \end{subfigure}
    \begin{subfigure}[c]{0.15\textwidth}
        \centering
        \includegraphics[width=\linewidth]{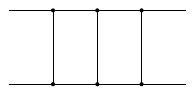}
        \caption{}\label{fig5b}
    \end{subfigure}
    \begin{subfigure}[c]{0.15\textwidth}
        \centering
        \includegraphics[width=\linewidth]{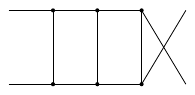}
        \caption{}\label{fig5c}
    \end{subfigure}

    \medskip

    \begin{subfigure}[c]{0.15\textwidth}
        \centering
        \includegraphics[width=\linewidth]{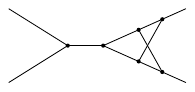}
        \caption{}\label{fig5d}
    \end{subfigure}
        \caption{Topologies associated with each integral family class. } \label{Fig:democlass}
        \centering
    \end{figure}

Here, we present the integral families that are encountered in the two-loop mixed QCD-EW corrections to $e^+e^-\rightarrow b \bar b$. Following the same notation as in~\cite{Armadillo:2022bgm}, we introduce the following symbols to represent different combinations of the loop and external momenta:
\begin{equation}
\begin{split}
&\mathcal{D}_i=q_i^2,~\mathcal{D}_{ij}=(q_i-q_j)^2,~\mathcal{D}_{i;j}=(q_i-p_j)^2,
    \\
    &\mathcal{D}_{i;jk}=(q_i-p_j-p_k)^2,~\mathcal{D}_{ij;k}=(q_i-q_j-p_k)^2,
    \\
    &\mathcal{D}_{ij;kl}=(q_i-p_j-p_k-p_l)^2,
    \\
    &\mathcal{D}_{ij;kl,m}=(q_i-q_j-p_k-p_l+p_m)^2.
\end{split}
\end{equation}
In terms of these symbols, a sample Feynman integral reads:
\begin{equation}
\begin{split}
    I=\int\frac{d^dq_1}{(2\pi)^d}\frac{d^dq_2}{(2\pi)^d}\frac{1}{\mathcal{D}_1^{\alpha_1} \mathcal{D}_2^{\alpha_2}(\mathcal{D}_{12}-m_1^2)^{\alpha_3}...}.
\end{split}
\end{equation}
We separate the integral families into four classes, $A$, $B$, $C$, and $D$, according to the combinations of momenta that appear in the propagator denominators.  
The topologies associated with each class are presented in Figure~\ref{Fig:democlass}, while the details of the integral families that belong to each class are provided below.

The $A$-class integral families are:
\begin{align}
    &A_1:\{\mathcal{D}_1,\mathcal{D}_2-m_1^2,\mathcal{D}_{12}-m_1^2,\mathcal{D}_{1;1},\mathcal{D}_{2;1}, \nonumber
    \\
    &~~~~~~~~~~~~\mathcal{D}_{1;12},\mathcal{D}_{12;2},\mathcal{D}_{1;3}-m_1^2,\mathcal{D}_{2,3}\} \nonumber \displaybreak[0]
    \\
    &A_2:\{\mathcal{D}_1,\mathcal{D}_2-m_1^2,\mathcal{D}_{12}-m_1^2,\mathcal{D}_{1;1},\mathcal{D}_{2;1}, \nonumber
    \\
    &~~~~~~~~~~~~\mathcal{D}_{1;12}-m_2^2,\mathcal{D}_{12;2},\mathcal{D}_{1;3}-m_1^2,\mathcal{D}_{2,3}\} \nonumber \displaybreak[0]
    \\
    &A_3:\{\mathcal{D}_1-m_1^2,\mathcal{D}_2,\mathcal{D}_{12}-m_1^2,\mathcal{D}_{1;1}, \nonumber
    \\
    &~~~~~~~~~~~~~~~~~~~\mathcal{D}_{2;1},\mathcal{D}_{1;12}-m_1^2,\mathcal{D}_{12;2},\mathcal{D}_{1;3},\mathcal{D}_{2,3}-m_1^2\} \nonumber
    \\
    &A_4:\{\mathcal{D}_1-m_1^2,\mathcal{D}_2-m_2^2,\mathcal{D}_{12}-m_2^2,\mathcal{D}_{1;1},\mathcal{D}_{2;1}, \nonumber
    \\
    &~~~~~~~~~~~~~~~~~~~\mathcal{D}_{1;12},\mathcal{D}_{12;2},\mathcal{D}_{1;3}-m_2^2,\mathcal{D}_{2,3}\} \nonumber \displaybreak[0]
    \\
    &A_5:\{\mathcal{D}_1-m_1^2,\mathcal{D}_2,\mathcal{D}_{12}-m_2^2,\mathcal{D}_{1;1},\mathcal{D}_{2;1}, \nonumber
    \\
    &~~~~~~~~~~~~~~~~~~~\mathcal{D}_{1;12}-m_1^2,\mathcal{D}_{12;2},\mathcal{D}_{1;3}-m_2^2,\mathcal{D}_{2,3}\} \nonumber \displaybreak[0]
    \\
    &A_6:\{\mathcal{D}_1-m_1^2,\mathcal{D}_2-m_2^2,\mathcal{D}_{12}-m_2^2,\mathcal{D}_{1;1},\mathcal{D}_{2;1}, \nonumber
    \\
    &~~~~~~~~~~~~~~~~~~~~\mathcal{D}_{1;12}-m_1^2,\mathcal{D}_{12;2},\mathcal{D}_{1;3}-m_2^2,\mathcal{D}_{2,3}\} \nonumber \displaybreak[0]
    \\
    &A_7:\{\mathcal{D}_1-m_1^2,\mathcal{D}_2-m_2^2,\mathcal{D}_{12}-m_1^2,\mathcal{D}_{1;1},\mathcal{D}_{2;1}, \nonumber
    \\
    &~~~~~~~~~~~~~~~~~~~\mathcal{D}_{1;12}-m_1^2,\mathcal{D}_{12;2},\mathcal{D}_{1;3},\mathcal{D}_{2,3}-m_1^2\} \nonumber \displaybreak[0]
    \\
    &A_8:\{\mathcal{D}_1-m_1^2,\mathcal{D}_2,\mathcal{D}_{12}-m_1^2,\mathcal{D}_{1;1},\mathcal{D}_{2;1}, \nonumber
    \\
    &~~~~~~~~~~~~~~~~~~~\mathcal{D}_{1;12}-m_1^2,\mathcal{D}_{12;2},\mathcal{D}_{1;3}-m_2^2,\mathcal{D}_{2,3}-m_1^2\} \nonumber \displaybreak[0]
    \\
    &A_9:\{\mathcal{D}_1-m_1^2,\mathcal{D}_2-m_2^2,\mathcal{D}_{12}-m_2^2,\mathcal{D}_{1;1},\mathcal{D}_{2;1}, \nonumber
    \\
    &~~~~~~~~~~~~~~~~~~~\mathcal{D}_{1;12}-m_3^2,\mathcal{D}_{12;2},\mathcal{D}_{1;3}-m_2^2,\mathcal{D}_{2,3}-m_1^2\} \nonumber \displaybreak[0]
    \\
    &A_{10}:\{\mathcal{D}_1-m_1^2,\mathcal{D}_2-m_2^2,\mathcal{D}_{12}-m_3^2,\mathcal{D}_{1;1},\mathcal{D}_{2;1}, \nonumber
    \\
    &~~~~~~~~~~~~~~~~~~~~\mathcal{D}_{1;12}-m_1^2,\mathcal{D}_{12;2},\mathcal{D}_{1;3}-m_3^2,\mathcal{D}_{2,3}\} \nonumber \displaybreak[0]
    \\
    &A_{11}:\{\mathcal{D}_1-m_1^2,\mathcal{D}_2-m_2^2,\mathcal{D}_{12}-m_3^2,\mathcal{D}_{1;1},\mathcal{D}_{2;1}, \nonumber
    \\
    &~~~~~~~~~~~~~~~~~~~~\mathcal{D}_{1;12}-m_1^2,\mathcal{D}_{12;2},\mathcal{D}_{1;3},\mathcal{D}_{2,3}-m_3^2\} \nonumber \displaybreak[0]
    \\
    &A_{12}:\{\mathcal{D}_1-m_1^2,\mathcal{D}_2,\mathcal{D}_{12}-m_1^2,\mathcal{D}_{1;1},\mathcal{D}_{2;1}, \nonumber
    \\
    &~~~~~~~~~~~~~~~~~~~~\mathcal{D}_{1;12}-m_1^2,\mathcal{D}_{12;2},\mathcal{D}_{1;3}-m_2^2,\mathcal{D}_{2,3}-m_3^2\}  \displaybreak[0]
\end{align}
The $B$-class integral families are:
\begin{align} 
    &B_1:\{\mathcal{D}_1,\mathcal{D}_2,\mathcal{D}_{12},\mathcal{D}_{1;1},\mathcal{D}_{2;1}, \nonumber
    \\
    &~~~~~~~~~~~~\mathcal{D}_{1;12},\mathcal{D}_{2;12},\mathcal{D}_{1;3},\mathcal{D}_{2,3}\} \nonumber \displaybreak[0]
    \\
    &B_2:\{\mathcal{D}_1-m_1^2,\mathcal{D}_2,\mathcal{D}_{12},\mathcal{D}_{1;1},\mathcal{D}_{2;1}, \nonumber
    \\
    &~~~~~~~~~~~~~~~~~~~\mathcal{D}_{1;12}-m_1^2,\mathcal{D}_{2;12},\mathcal{D}_{1;3},\mathcal{D}_{2,3}\} \nonumber \displaybreak[0]
    \\
    &B_3:\{\mathcal{D}_1-m_1^2,\mathcal{D}_2,\mathcal{D}_{12}-m_1^2,\mathcal{D}_{1;1},\mathcal{D}_{2;1}, \nonumber
    \\
    &~~~~~~~~~~~~~~~~~~~\mathcal{D}_{1;12}-m_1^2,\mathcal{D}_{2;12},\mathcal{D}_{1;3},\mathcal{D}_{2,3}\} \nonumber \displaybreak[0]
    \\
    &B_4:\{\mathcal{D}_1-m_1^2,\mathcal{D}_2,\mathcal{D}_{12},\mathcal{D}_{1;1},\mathcal{D}_{2;1}-m_2^2, \nonumber
    \\
    &~~~~~~~~~~~~~~~~~~~\mathcal{D}_{1;12}-m_1^2,\mathcal{D}_{2;12},\mathcal{D}_{1;3},\mathcal{D}_{2,3}\} \nonumber \displaybreak[0]
    \\
    &B_5:\{\mathcal{D}_1-m_1^2,\mathcal{D}_2-m_1^2,\mathcal{D}_{12},\mathcal{D}_{1;1},\mathcal{D}_{2;1}, \nonumber
    \\
    &~~~~~~~~~~~~~~~~~~~\mathcal{D}_{1;12}-m_1^2,\mathcal{D}_{2;12}-m_1^2,\mathcal{D}_{1;3},\mathcal{D}_{2,3}\} \nonumber \displaybreak[0]
    \\
    &B_6:\{\mathcal{D}_1-m_1^2,\mathcal{D}_2-m_1^2,\mathcal{D}_{12}-m_2^2,\mathcal{D}_{1;1},\mathcal{D}_{2;1}, \nonumber
    \\
    &~~~~~~~~~~~~~~~~~~~\mathcal{D}_{1;12}-m_1^2,\mathcal{D}_{2;12}-m_1^2,\mathcal{D}_{1;3},\mathcal{D}_{2,3}\} \nonumber \displaybreak[0]
    \\
    &B_7:\{\mathcal{D}_1-m_1^2,\mathcal{D}_2-m_2^2,\mathcal{D}_{12},\mathcal{D}_{1;1},\mathcal{D}_{2;1}, \nonumber
    \\
    &~~~~~~~~~~~~~~~~~~~\mathcal{D}_{1;12}-m_1^2,\mathcal{D}_{2;12}-m_2^2,\mathcal{D}_{1;3},\mathcal{D}_{2,3}\} \nonumber \displaybreak[0]
    \\
    &B_8:\{\mathcal{D}_1,\mathcal{D}_2-m_1^2,\mathcal{D}_{12}-m_1^2,\mathcal{D}_{1;1},\mathcal{D}_{2;1}, \nonumber
    \\
    &~~~~~~~~~~~~\mathcal{D}_{1;12}-m_2^2,\mathcal{D}_{2;12}-m_1^2,\mathcal{D}_{1;3},\mathcal{D}_{2,3}\} \nonumber \displaybreak[0]
    \\
    &B_9:\{\mathcal{D}_1-m_1^2,\mathcal{D}_2-m_2^2,\mathcal{D}_{12}-m_1^2,\mathcal{D}_{1;1},\mathcal{D}_{2;1}, \nonumber
    \\
    &~~~~~~~~~~~~~~~~~~~\mathcal{D}_{1;12}-m_1^2,\mathcal{D}_{2;12}-m_2^2,\mathcal{D}_{1;3},\mathcal{D}_{2,3}\} \nonumber \displaybreak[0]
    \\
    &B_{10}:\{\mathcal{D}_1-m_1^2,\mathcal{D}_2-m_1^2,\mathcal{D}_{12},\mathcal{D}_{1;1},\mathcal{D}_{2;1}, \nonumber
    \\
    &~~~~~~~~~~~~~~~~~~~~~\mathcal{D}_{1;12}-m_1^2,\mathcal{D}_{2;12}-m_1^2,\mathcal{D}_{1;3}-m_2^2,\mathcal{D}_{2,3}\} \nonumber \displaybreak[0]
    \\
    &B_{11}:\{\mathcal{D}_1-m_1^2,\mathcal{D}_2-m_2^2,\mathcal{D}_{12}-m_3^2,\mathcal{D}_{1;1},\mathcal{D}_{2;1}, \nonumber
    \\
    &~~~~~~~~~~~~~~~~~~~~~\mathcal{D}_{1;12}-m_1^2,\mathcal{D}_{2;12}-m_2^2,\mathcal{D}_{1;3},\mathcal{D}_{2,3}\} \nonumber \displaybreak[0]
    \\
    &B_{12}:\{\mathcal{D}_1-m_1^2,\mathcal{D}_2-m_2^2,\mathcal{D}_{12},\mathcal{D}_{1;1},\mathcal{D}_{2;1}, \nonumber
    \\
    &~~~~~~~~~~~~~~~~~~~~~\mathcal{D}_{1;12}-m_3^2,\mathcal{D}_{2;12}-m_2^2,\mathcal{D}_{1;3},\mathcal{D}_{2,3}\} \nonumber \displaybreak[0]
    \\
    &B_{13}:\{\mathcal{D}_1-m_1^2,\mathcal{D}_2-m_2^2,\mathcal{D}_{12}-m_2^2,\mathcal{D}_{1;1},\mathcal{D}_{2;1}, \nonumber
    \\
    &~~~~~~~~~~~~~~~~~~~~~\mathcal{D}_{1;12}-m_3^2,\mathcal{D}_{2;12}-m_2^2,\mathcal{D}_{1;3},\mathcal{D}_{2,3}\}  \displaybreak[0]
\end{align}
The $C$-class integral families are:
\begin{align}
    &C_1:\{\mathcal{D}_1-m_1^2,\mathcal{D}_2,\mathcal{D}_{12}-m_1^2,\mathcal{D}_{1;2},\mathcal{D}_{2;2}, \nonumber
    \\
    &~~~~~~~~~~~~\mathcal{D}_{1;12}-m_1^2,\mathcal{D}_{2;12},\mathcal{D}_{1;3},\mathcal{D}_{2,3}\} \nonumber \displaybreak[0]
    \\
    &C_2:\{\mathcal{D}_1,\mathcal{D}_2,\mathcal{D}_{12}-m_1^2,\mathcal{D}_{1;2},\mathcal{D}_{2;2}, \nonumber
    \\
    &~~~~~~~~~~~~\mathcal{D}_{1;12},\mathcal{D}_{2;12}-m_1^2,\mathcal{D}_{1;3}-m_1^2,\mathcal{D}_{2,3}\} \nonumber \displaybreak[0]
    \\
    &C_3:\{\mathcal{D}_1,\mathcal{D}_2-m_1^2,\mathcal{D}_{12}-m_1^2,\mathcal{D}_{1;2},\mathcal{D}_{2;2}, \nonumber
    \\
    &~~~~~~~~~~~~\mathcal{D}_{1;12}-m_2^2,\mathcal{D}_{2;12}-m_1^2,\mathcal{D}_{1;3},\mathcal{D}_{2,3}\} \nonumber \displaybreak[0]
    \\
    &C_4:\{\mathcal{D}_1-m_1^2,\mathcal{D}_2-m_2^2,\mathcal{D}_{12}-m_1^2,\mathcal{D}_{1;2},\mathcal{D}_{2;2}, \nonumber
    \\
    &~~~~~~~~~~~~~~~~~~~~\mathcal{D}_{1;12}-m_1^2,\mathcal{D}_{2;12}-m_2^2,\mathcal{D}_{1;3},\mathcal{D}_{2,3}\} \nonumber \displaybreak[0]
    \\
    &C_5:\{\mathcal{D}_1,\mathcal{D}_2,\mathcal{D}_{12}-m_1^2,\mathcal{D}_{1;2},\mathcal{D}_{2;2}, \nonumber
    \\
    &~~~~~~~~~~~~\mathcal{D}_{1;12}-m_2^2,\mathcal{D}_{2;12}-m_1^2,\mathcal{D}_{1;3}-m_1^2,\mathcal{D}_{2,3}\} \nonumber \displaybreak[0]
    \\
    &C_6:\{\mathcal{D}_1-m_1^2,\mathcal{D}_2,\mathcal{D}_{12}-m_2^2,\mathcal{D}_{1;2},\mathcal{D}_{2;2}, \nonumber
    \\
    &~~~~~~~~~~~~~~~~~~~~\mathcal{D}_{1;12},\mathcal{D}_{2;12}-m_2^2,\mathcal{D}_{1;3}-m_2^2,\mathcal{D}_{2,3}\} \nonumber \displaybreak[0]
    \\
    &C_7:\{\mathcal{D}_1-m_1^2,\mathcal{D}_2,\mathcal{D}_{12}-m_2^2,\mathcal{D}_{1;2},\mathcal{D}_{2;2}, \nonumber
    \\
    &~~~~~~~~~~~~~~~~~~~~\mathcal{D}_{1;12}-m_1^2,\mathcal{D}_{2;12}-m_2^2,\mathcal{D}_{1;3}-m_2^2,\mathcal{D}_{2,3}\}  \displaybreak[0]
\end{align}
The $D$-class integral families are:
\begin{align}
    &D_1:\{\mathcal{D}_1-m_1^2,\mathcal{D}_2-m_1^2,\mathcal{D}_{1;1},\mathcal{D}_{1;12},\mathcal{D}_{12}, \nonumber
    \\
    &~~~~~~~~~~~~\mathcal{D}_{12:1},\mathcal{D}_{12;12},\mathcal{D}_{12;12,3}-m_1^2,\mathcal{D}_{2,3}\} \nonumber \displaybreak[0]
    \\
    &D_2:\{\mathcal{D}_1-m_1^2,\mathcal{D}_2-m_1^2,\mathcal{D}_{1;1},\mathcal{D}_{1;12}-m_1^2,\mathcal{D}_{12}-m_2^2, \nonumber
    \\
    &~~~~~~~~~~~~~~~~~~~~\mathcal{D}_{12:1},\mathcal{D}_{12;12},\mathcal{D}_{12;12,3}-m_1^2,\mathcal{D}_{2,3}\} \nonumber \displaybreak[0]
    \\
    &D_3:\{\mathcal{D}_1-m_1^2,\mathcal{D}_2-m_2^2,\mathcal{D}_{1;1},\mathcal{D}_{1;12}-m_1^2,\mathcal{D}_{12}-m_3^2, \nonumber
    \\
    &~~~~~~~~~~~~~~~~~~~~\mathcal{D}_{12:1},\mathcal{D}_{12;12},\mathcal{D}_{12;12,3}-m_1^2,\mathcal{D}_{2,3}\}  \displaybreak[0]
\end{align}

\section{Cluster Information Table}\label{appendix_cluster}

We provide the cluster information that is used to run the workflow of this project in the table below:

\phantom{a}
\begin{table}[ht!]
\centering
\label{tab:clusters}
\renewcommand{\arraystretch}{1.3}
\begin{tabular}{@{}llcl@{}}
\Xhline{1pt}
Cluster        & CPU                            & Cores & Memory        \\
\hline
INPAC bl-5   & 2$\times$ AMD EPYC 7763        & 128   & 512\,GB DDR4  \\
XINCHEN        & 2$\times$ Intel Xeon E5-2698 v3 & 32   & 512\,GB DDR4  \\
PCEPC          & 2$\times$ AMD EPYC 7763        & 128   & 3072\,GB DDR4 \\
\Xhline{1pt}
\end{tabular}
\caption{Hardware specifications of the computing clusters used in this work.}
\end{table}

\bibliography{long}
\bibliographystyle{JHEP}

\end{document}